\documentclass[12pt]{article}
\usepackage{epsfig,amsfonts,amssymb}
\begin{document}

\voffset -0.7 true cm
\hoffset 1.1 true cm
\topmargin 0.0in
\evensidemargin 0.0in
\oddsidemargin 0.0in
\textheight 8.6in
\textwidth 7.1in
\parskip 10 pt
 
\def\Tr{\hbox{Tr}}
\newcommand{\be}{\begin{equation}}
\newcommand{\ee}{\end{equation}}
\newcommand{\bea}{\begin{eqnarray}}
\newcommand{\eea}{\end{eqnarray}}
\newcommand{\beas}{\begin{eqnarray*}}
\newcommand{\eeas}{\end{eqnarray*}}
\def\kl{{\frac{2 \pi l}{\beta}}}
\def\km{{\frac{2 \pi m}{\beta}}}
\def\kn{{\frac{2 \pi n}{\beta}}}
\def\kr{{\frac{2 \pi r}{\beta}}}
\def\ks{{\frac{2 \pi s}{\beta}}}
\def\b{{\beta}}
\font\cmsss=cmss8
\def\C{{\hbox{\cmsss C}}}
\font\cmss=cmss10
\def\bigC{{\hbox{\cmss C}}}
\def\scriptlap{{\kern1pt\vbox{\hrule height 0.8pt\hbox{\vrule width 0.8pt
  \hskip2pt\vbox{\vskip 4pt}\hskip 2pt\vrule width 0.4pt}\hrule height 0.4pt}
  \kern1pt}}
\def\ba{{\bar{a}}}
\def\bb{{\bar{b}}}
\def\bc{{\bar{c}}}
\def\bphi{{\Phi}}
\def\Bigggl{\mathopen\Biggg}
\def\Bigggr{\mathclose\Biggg}
\def\Biggg#1{{\hbox{$\left#1\vbox to 25pt{}\right.\n@space$}}}
\def\n@space{\nulldelimiterspace=0pt \m@th}
\def\m@th{\mathsurround = 0pt}

\begin{titlepage}
\begin{flushright}
{\small TIFR/TH/04-05} \\
{\small hep-th/0403203}
\end{flushright}

\begin{center}

\vspace{2mm}

{\Large \bf An Inflationary Model in String Theory}

\vspace{3mm}

Norihiro Iizuka${}^1$ and Sandip P. Trivedi${}^2$

\vspace{1mm}

{\small \sl Tata Institute of Fundamental Research} \\
{\small \sl Homi Bhabha Road, Mumbai, 400 005, INDIA} \\
{\small \tt iizuka@theory.tifr.res.in${}^1$, sandip@tifr.res.in${}^2$}
\vspace{1mm}

\end{center}

\vskip 0.3 cm

\noindent
We construct a model of inflation in string theory 
after carefully taking into account moduli stabilization. 
The setting is a warped compactification of Type 
IIB string theory in the presence of D3 and anti-D3-branes. The inflaton is the position of a 
D3-brane in the internal space. By suitably adjusting fluxes and the location of symmetrically placed anti-D3-branes, 
we show that at a point of enhanced symmetry, the inflaton potential $V$
can have a broad maximum, satisfying the condition $V^{''}/V \ll 1$ in 
Planck units. On starting close to the top of this potential the slow-roll 
conditions can be met. Observational constraints impose significant restrictions.
As a first pass we show that these can be satisfied and determine the 
important scales in the compactification to within an order of magnitude. 
One robust feature is that the scale of inflation is low, $H={\cal{O}}(10^{10})$ GeV.
Removing the observational constraints makes it much easier to 
construct a slow-roll inflationary model. 
Generalizations and consequences including the possibility of 
eternal inflation are also discussed. 
A more careful study, including explicit constructions of 
the model in string theory, is left for the future.

\end{titlepage}

%%%%%%%%%%%%%%%%%%%%%%%%%%%%%%%%%%%%%%%%%%%%%%%%%%%%%%%%%%%%%%%%%%%%%%%%%%%%%%%%%%%%%%%%%%%%%%%%%%%%
\section{Introduction}
%%%%%%%%%%%%%%%%%%%%%%%%%%%%%%%%%%%%%%%%%%%%%%%%%%%%%%%%%%%%%%%%%%%%%%%%%%%%%%%%%%%%%%%%%%%%%%%%%%%%

Inflation is an attractive idea that solves many important problems in cosmology.
It is also in good agreement with current observational data.
It is therefore important to understand if inflation can arise in string theory.

Despite several attempts no satisfactory model of inflation in string theory 
has been constructed as yet. This problem is closely tied to the issue of
moduli stabilization.
It is well known that a very flat potential is required for inflation.
There are many light fields, called moduli, in string theory which to first approximation have an exactly
 flat potential. So it might seem at first that a slowly enough varying potential can be easily generated.
However, any attempt to generate such a potential typically runs into difficulty.
One finds that some very unstable direction develops, along which the potential descends
much too rapidly. As a result, the required flatness condition is
not met and inflation is not realized.
Recently, there has been considerable progress in moduli stabilization.
So one can be hopeful that these obstacles will be overcome.

In this paper we outline the construction of a model for inflation in string theory.
Our construction is based on the recent developments in moduli stabilization. 
The setting is warped flux compactifications of type IIB string theory, in the presence of D3-branes
and anti-D3-branes.
Using fluxes we fix all the complex structure moduli of the Calabi-Yau space
and the dilaton-axion \cite{GKP}.
We also use a generic superpotential of the kind which arises due to gaugino condensation
to stabilize the volume modulus \cite{KKLT} (KKLT).
Our discussion of inflation is closely related to the recent attempt in \cite{KKLMMT} (KKLMMT). 
In particular the inflaton in our model is a D3-brane modulus, i.e.,
a scalar field which corresponds to the location of the D3-brane in the internal Calabi-Yau space.
In \cite{KKLMMT} it was argued that a superpotential of the kind mentioned above
leads to the D3-brane moduli acquiring a mass, $m^2 =1/6 R$, where $R$ is the curvature scalar.
This mass is too big and ruins the required flatness of the
potential.

We explore a small twist on this story here.
Consider a Calabi-Yau space with a $Z_2$ symmetry\footnote{This $Z_2$ symmetry need not be the one
involved in the orientifold action.}.
A mobile D3-brane is located in the vicinity of the $Z_2$ symmetric point 
and experiences an attractive Coulomb force due to two symmetrically located anti-D3-branes.
In such a situation we show that by adjusting the fluxes and the brane-anti-brane separation, 
the Coulomb attraction can nearly cancel the effect of the curvature induced mass mentioned above.
As a result the $Z_2$ symmetric point turns into a maximum of the potential. The near cancellation results in a 
broad maximum, with $|m^2|/H^2 \ll 1$, where $H$, is the Hubble scale corresponding to the
height of the potential at the maximum.
By starting close enough to such a  maximum the required
conditions for slow-roll inflation can then all be met.

The observational constraints, especially  the scale of  density perturbations, impose  stringent
restrictions on the model. 
We show that the constraints can be met and  determine the important scales in 
the compactification to within an order of magnitude\footnote{More accurately, our estimates of some of the
 important energy   scales involved is uncertain by factors of order unity. Up to these 
uncertainties we show 
that the constraints can be met.}.    
A robust feature of our model, independent of many details,  is that the  scale of inflation is low.
The Hubble scale, $H$, is of order $10^{10}$ GeV, which corresponds to a cosmological constant of order
$10^{14}$ GeV. Thus the production of tensor perturbations is highly suppressed.
The observation of gravity waves by the Planck experiment would therefore rule out this model.

A more careful study of whether all the constraints can be met  will require  concrete constructions of the
 model in string theory and   is left for the future. The non-perturbative superpotential we evoke, and the 
assumption that the full potential can be obtained by adding 
the brane-anti-brane interaction to the term coming from
the superpotential\footnote{This assumption would be correct if the brane-anti-brane interaction arise from 
a $D$-term. Evidence in support of this has been found in \cite{Dvali:2003zh,Binetruy:2004hh}. We thank S. Kachru for 
bringing these references to our attention.} also needs to be studied further. 
The last two issues are common to many KKLT type constructions.

As a model for inflation our construction is incomplete in three ways. We have not addressed how inflation ends,
how it begins, and how the standard model can be incorporated in it. 
Ending inflation successfully requires adequate reheating. This  depends on how the standard model is 
incorporated. In terms of beginning inflation, it could be that the model does not depend sensitively on 
initial conditions. 
It has been argued that a broad maximum of the kind in this model gives rise to eternal inflation.
Regions where quantum fluctuations have driven the inflaton to the top of the potential hill
grow exponentially more rapidly and soon dominate the universe, regardless of initial conditions.
The inflationary epoch discussed in this paper then arises when fluctuations cause 
the scalar field to descend far enough from the top so that the classical evolution becomes dominant. 
This is an appealing picture but it needs to be understood better. 
We leave these issues for the future.

The important features of this model are quite general.
They  essentially depend only   on the existence of a broad maximum,
with $|m|^2/H^2 \ll 1$, and are independent of most details. For example, we have emphasized the role
of a $Z_2$ symmetry above. But the idea works more generally, even when there is no such symmetry, 
for a D3-brane located between  two appropriately positioned anti-D3-branes.
More generally this construction can be viewed as an existence proof for broad maxima in the landscape of 
string theory. It seems reasonable to believe that there are many such maxima, with the complex structure moduli
or K\"ahler moduli also playing the role of the inflaton\footnote{S.P.T. thanks M. Douglas for emphasizing this
point.}.
The inflationary parameters, like the scale of density perturbations, or the tilt in the
spectrum of scalar perturbations, probably take many different values at these maxima, most of 
which will not agree with observation.
Further progress in moduli stabilization will allow us to test this grim possibility.

This paper is organized as follows.
Our  basic set-up is discussed in section 2.1. The positive mass for the brane moduli
due to the curvature coupling, is reviewed in section 2.2. In section 2.3, we discuss the  potential for a
D3-brane located between two  symmetrically placed anti-D3-branes, 
and show that a broad maximum can arise. The resulting inflationary
scenario is discussed in section 2.4. The constraints on the compactification which arise are analyzed
in section 2.5. 
We close with an extended discussion in section 3.

Before proceeding we should comment on some of the relevant literature.
The idea of brane inflation was first discussed by \cite{DT}.
Other related papers are  \cite{others}, \cite{angles}.
In particular, while not worrying about moduli stabilization, 
\cite{BMNQ} showed that a flat potential could be obtained by considered symmetrically positioned
 branes.
For a review, see \cite{Quevedo}. 
Standard textbook references for inflation are \cite{Linde} and 
\cite{Liddle}, see also \cite{LythRiotto}. Some relevant references for moduli stabilization
are \cite{GVW} and \cite{GKP} for general framework, and \cite{KSTetal} for some  examples.
For recent progress towards meeting the conditions of the KKLT construction, see \cite{GKTT}.
\cite{EGQ} explores the KKLT construction further, \cite{BKQt} considers inducing the anti-D3-brane 
charge on D7-branes.  
A variant of the KKLT scenario which does not require the anti-D3-brane 
is \cite{Saltman:2004sn}. An investigation of de Sitter vacua using F-term potentials and additional
light moduli is in \cite{Brustein:2004xn}. 
A recent attempt to overcome the problems faced in KKLMMT
involves the use of a shift symmetry \cite{shiftsymmetry}. It would be nice to see if shift symmetry
is present in Calabi-Yau orientifolds or their F-theory generalizations,
which are required for controlled stabilization of all moduli and preserve only ${\cal N}=1$ 
supersymmetry. For an  attempt in the context of string theory 
to use higher derivative terms for inflation see  \cite{ST}. 
Inflation  with a quadratic potential of the kind we obtain here was studied earlier in 
\cite{German:2001tz}, which  arrived at similar conclusions  about the low energy scale during 
inflation etc\footnote{We thank S. Sarkar for bringing this paper to our attention.}. 
Two related papers on brane inflation appeared  while our manuscript was being readied.
\cite{Burgess:2004kv} uses additional D-terms obtained by adding the 
Standard Model fields in the KKLMMT set up and argues numerically that inflation can be obtained,
and \cite{DeWolfe:2004qx} explores the possibility of inflation in the dynamics of more than one 
anti-brane in a K-S throat.

Our conventions are as follows. $M_{10}$  refers to  the ten-dimensional Planck scale.
It is related to the string scale, $\alpha^{'}$, and the string coupling, $g_s$, by 
${1 \over M_{10}^8}= {1\over 2}(2\pi)^7(\alpha^{'})^{4}g_s^2.$
$M_{Pl}$  refers to  the four-dimensional Planck scale. It is  defined by 
$M_{Pl}^2={1 \over 8 \pi G_N}$ and satisfies the relation,  $M_{Pl}^2=M_{10}^8 L^6$,
where $L^6$ is the volume of the six-dimensional internal space. Finally,
the tension of the D3-brane is given by $T_3={1\over (2\pi)^3 (\alpha^{'})^2 g_s}$.

%%%%%%%%%%%%%%%%%%%%%%%%%%%%%%%%%%%%%%%%%%%%%%%%%%%%%%%%%%%%%%%%%%%%%%%%%%%%%%%%%%%%%%%%%%%%%%%%%%%%
\section{The Model}
%%%%%%%%%%%%%%%%%%%%%%%%%%%%%%%%%%%%%%%%%%%%%%%%%%%%%%%%%%%%%%%%%%%%%%%%%%%%%%%%%%%%%%%%%%%%%%%%%%%%
%%%%%%%%%%%%%%%%%%%%%%%%%%%%%%%%%%%%%%%%%%%%%%%%%%%%%%%%%%%%%%%%%%%%%%%%%%%%%%%%%%%%%%%%%%%%%%%%%%%%
\subsection{Basic Set-up}
%%%%%%%%%%%%%%%%%%%%%%%%%%%%%%%%%%%%%%%%%%%%%%%%%%%%%%%%%%%%%%%%%%%%%%%%%%%%%%%%%%%%%%%%%%%%%%%%%%%%

Consider  IIB string theory on a six-dimensional Calabi-Yau orientifold,
with the three forms $H_3,F_3$ turned on. 
More generally we can consider F-theory on an elliptically fibered Calabi-Yau fourfold. 
The resulting compactification is of the warped kind, \cite{GKP}, 
\be
\label{warpmet}
ds^2= e^{2A(y)} \eta_{\mu\nu} dx^\mu dx^\nu + e^{-2A(y)} g_{mn} dy^m dy^n \,.
\ee
The five-form $F_5$ is non-zero,
and determined. The three-forms give rise to a superpotential for the complex structure moduli,
\cite{GVW},
\be
\label{superpotgvw}
W=\int G_{3} \wedge \Omega \,,
\ee
where $G_{3} = F_{3} - \tau H_{3}$, $\tau$ is the dilaton axion field, and 
$\Omega$ is the homomorphic three-form on the Calabi-Yau space. This superpotential
in general fixes all the complex structure moduli and the dilaton-axion.

As was discussed in \cite{GKP}, such a construction can provide a compactification of the 
Klebanov-Strassler (K-S)
deformed conifold solution \cite{KS}. By tunning the fluxes the complex structure 
moduli can be stabilized close to a conifold singularity. 
An intuitive picture of the resulting compactification is as follows. 
Roughly speaking, the compactification contains a small three-sphere threaded by flux.
The resulting backreaction is significant and causes a ``throat" to develop - this is a region
where the warp factor, $e^{2A(y)}$, departs significantly from unity. Unlike in the case of $AdS_5$, the 
K-S throat terminates on a three-sphere where the warp factor acquires its minimum value. 
It is relevant to note for our purposes that if the Calabi-Yau manifold has discrete symmetries, 
more than one small three-sphere can be present
when the complex structure moduli are stabilized close to the conifold point. These three-spheres 
would be symmetrically 
located about a point of enhanced symmetry  and in turn would give rise to symmetrically located throat regions
where the warp factor departs significantly from unity. 

In the subsequent discussion we restrict ourselves to Calabi-Yau orientifolds with one K\"ahler modulus,
the volume. As discussed in \cite{Witten:1996bn}, non-perturbative corrections to the superpotential, 
for example due to gaugino condensation on wrapped D7-branes, can arise.
These are dependent on the volume and can stabilize it \cite{KKLT}. 
Additional anti-D3-branes at the bottom of one (or more) K-S throats
can lift these vacua to positive cosmological constant giving rise to dS space. 

Finally, mobile D3-branes can be present in the compactification. 
Their interaction with anti-D3-branes can be calculated.
The idea explored in \cite{KKLMMT} was that the attractive potential between a mobile
brane and an anti-brane might give rise to a slowly varying potential suitable for inflation.
However, a detailed analysis of the resulting potential showed that when the details of
volume stabilization, as mentioned above, are included, the D3-brane acquires a mass
which is too big to allow for the slow-roll conditions to be met. 
   
The new element we  consider in this paper   is to take  a Calabi-Yau space
with a $Z_2$  symmetry and  two symmetrically located K-S throats each containing an 
 anti-D3-brane. The mobile D3-brane is located in between in the vicinity of the $Z_2$ 
symmetric point.  We will see that in such a situation the positive mass term due to the 
curvature coupling, can be canceled to good approximation by the brane-anti-brane potential, 
giving rise to  a maximum in the potential energy with a  small mass, $|m|^2\ll H^2$.
By starting close to the maximum the  requirements for slow-roll inflation can  be met.

The rest of this section is organized as follows. 
We first briefly sketch out how the curvature coupling and related positive mass term arises
in \cite{KKLMMT}. Next we include the brane-anti-brane interaction and analyze the resulting
potential. A discussion of the resulting inflationary scenario follows in the section  2.4. 
The constraints imposed on the compactification are discussed in section 2.5.

\subsection{The Curvature Coupling: A positive mass}

It is useful to consider the dynamics of the mobile D3-brane in an effective theory 
obtained by integrating out the complex structure moduli.
This effective theory contains four complex scalar fields (with their fermionic partners     
to form chiral superfields).
$\bphi^i, i=1, \cdots, 3$, are three complex fields related to the D3-brane location.               
$\rho$ is an additional complex scalar, its real part is related to the volume $r$ by,
\be
\label{rela}
2r=\rho +{\bar \rho}-k(\bphi^i,{\bar \bphi^i}) \,.
\ee
(More correctly $r$ is proportional to the volume of the Calabi-Yau manifold, in the 
notation\footnote{$r$ in this paper is related to the field $\rho$, in \cite{GKP}, appendix A.1, after eq.~(A.2),
as follows:  $2r=-i\rho$.} of  \cite{GKP}, 
$r \sim e^{4u}$). $k(\bphi^i, {\bar \bphi^i})$ is the K\"ahler potential of the Calabi-Yau manifold.

The kinetic energy terms can be derived from the K\"ahler potential, \cite{DeWolfe:2002nn},
\be
\label{kp}
K=-3 \log(\rho +{\bar \rho}-k(\bphi^i,  {\bar \bphi^i})) \,.
\ee

The superpotential (in the absence of the anti-branes) takes the form 
\be
\label{superpot}
W=W_0+Ae^{-a\rho} \,,
\ee
where $W_0,A,a$ are constants in the effective theory. The first term above, $W_0$, arises  
by replacing the complex structure moduli with their vacuum expectation 
values in eq.~(\ref{superpotgvw}). The second term arises due to non-perturbative effects.
These could be, for example, due to gaugino condensation on D7-branes wrapping four-cycles in the
Calabi-Yau space, or due to Euclidean D3-brane instantons, \cite{Witten:1996bn}.
The prefactor $A$ also depends on the  expectation values of the complex structure moduli. 
The superpotential, eq.~(\ref{superpot}), 
 gives rise to a potential energy, 
\be
\label{potenr}
V^F={1 \over 6 r}\left(\partial_\rho W {\overline {\partial_\rho W}} (1+{1 \over 2r} k^{i{\bar j}} k_i k_{\bar j}) 
-{3 \over 2 r}({\overline W} \partial_\rho W + W \overline {\partial_\rho W}) \right) \,.
\ee

Including the effects of the anti-branes in this theory is subtle.
It can be shown (see appendix B of \cite{KKLMMT}) that the potential between a D3-brane and an anti-D3-brane
in a warped background takes the form\footnote{To relate this to eq.~(3.9) of \cite{KKLMMT}
at $\vec{r}=0$,
we note that the redshift  $Z^4$ is denoted by ${r_0^4 \over R^4}$ in \cite{KKLMMT}, and 
from the conventions discussed in the introduction above, it follows that 
${1 \over 2\pi^3}{T_3\over M_{10}^8}={R^4\over N} = 4 \pi g_s (\alpha^{'})^2$.} 
\be
\label{babpot}
V^B(\vec{r})=2T_3 Z^4 (1-{1 \over 2 \pi^3} {Z^4 T_3 \over M_{10}^8 |\vec{r} - \vec{r_1}|^4} ) \,.
\ee 
The first term on the r.h.s. is really the potential energy of the anti-brane. The second term 
arises due to the attractive RR and gravitational potential between the brane and anti-brane,
we will refer to it as the Coulomb term below.
$T_3$ is the tension of the $3$-brane, $\vec{r}, \vec{r_1}$ refer to the locations of the 
brane and anti-brane respectively, and  $Z^4=e^{4A(\vec{r_1})}$ refers to the redshift at 
the location of the anti-brane.

In eq.~(\ref{babpot}),  we have assumed that the metric in the background Calabi-Yau space
is flat, i.e., $g_{mn}=\delta_{mn}$, eq.~(\ref{warpmet}). More generally, in the the 
second term on the r.h.s, the factor $1/|\vec{r} - \vec{r_1}|^4$ will be replaced by the 
appropriate harmonic function. 
If more than one brane and anti-brane are present, this formula is generalized
in a straight forward manner. The first term on the r.h.s. includes the contributions of
 all the anti-branes with 
the appropriate redshift factor at their locations. The second term includes all 
brane-anti-brane pairs with the appropriate redshift factors and harmonic functions. 
We should also mention that the Coulomb term above is valid only when the brane-anti-brane 
separation is much bigger than the string scale, $|\vec{r}-\vec{r_1}| \gg \sqrt{\alpha^{'}}$,
and is much smaller than the size of the compactification, $L$, i.e., 
$|\vec{r}-\vec{r_1}| \ll L$. We will see in the following discussion that these conditions are 
indeed met\footnote{If $|\vec{r}-\vec{r_1}| \sim L$, the Harmonic function in eq.~(\ref{babpot})
 needs to be replaced by its compact space version.}. 

To include the effects of the anti-branes we will then simply assume that this potential due 
to the brane-anti-brane interactions can be added to $V^F$ above in  obtaining
 the full potential. 

The resulting total potential then takes the form,
\be
\label{totalpot}
V=V^F+V^B \,.
\ee

Keeping only the potential energy term of the anti-brane in eq.~(\ref{babpot}) and neglecting the 
Coulomb term for now, 
we get eq.~(5.14) of \cite{KKLMMT}, 
\be
\label{potb}
V={1 \over 6r}\left(\partial_\rho W \overline{\partial_\rho W}(1+{1 \over 2r} 
k^{i\bar{j}} k_i k_{\bar{j}} ) -{3 \over 2r}(\overline{W}\partial_\rho W+W 
\overline{\partial_\rho W})
\right) + {D \over (2r)^2} \,.
\ee
We should point out that this equation gives the potential in 
four-dimensional Planck units.  The second term above arises  from the anti-brane potential energy,
and is given by summing over all the anti-D3-branes, $2 T_3 \sum_i Z_i^4$. 
To convert to four-dimensional Planck units we use the relation $T_3/M_{Pl}^4 \sim 1/(2r)^3$.
Finally, we use the fact that the  redshift factor at the bottom of a K-S throat 
scales like $Z^4=e^{4A} \sim r$ for fixed integer fluxes. 
This gives the term ${D / (2r)^2}$ where the coefficient $D$ is independent of the volume.

In the vicinity of a point in moduli space where $k(\bphi^i, {\bar \bphi^i}) =\bphi^i {\bar \bphi^i}$
one can show that a dS minimum  exists  at $\rho=\rho_c, \bphi^i=
\bar{\bphi^j}=0$. The potential at the minimum is denoted by  $V=V_0(\rho_c)$.
Expanding around it the potential can be written as,
\be
\label{exppot}
V
={V_0(\rho_c)\over{\left(1-{\varphi\bar\varphi / 3M_{Pl}^2}\right)^2}}
 \approx
V_0(\rho_c)\left( 1+ {2\over 3} {\varphi \bar \varphi \over M_{Pl}^2}\right) \,.
\ee
Here $\varphi= \bphi\sqrt{3/(\rho+\bar{\rho})}$ is the canonically normalized field 
as follows from the K\"ahler potential (\ref{kp}), and we have inserted the  appropriate factor of $M_{Pl}$ 
required by dimensional analysis.
We see that the brane moduli acquire a positive mass. It is easy to see that 
$m^2={2 \over 3} {V_0(\rho_c) \over M_{Pl}^2}={1\over 6} R$, where $R$ is the curvature of the resulting 
dS space.

One more comment is in order before we proceed. 
The minima one gets from $V^F$ alone are supersymmetric and have negative cosmological 
constant. These are lifted to positive cosmological constant because of the 
anti-brane contribution, the $D/(2r)^2$ term in eq.~(\ref{potb}). 
At a typical dS minimum the contributions of the superpotential term and the anti-brane terms
in eq.~(\ref{potb}) are roughly comparable and each is of order $V_0({\rho_c})$.\footnote{The present 
day de Sitter 
phase would not be of this type, for this the two contributions would have to cancel to a very good accuracy 
leaving a small positive residual cosmological constant, but this is not the generic situation.}
This fact will be useful to bear in mind 
in the next subsection.

\subsection{Symmetrically Located Throats}

We are now ready to consider the new twist in this paper. 
Consider a situation mentioned above, where there are two K-S throats symmetrically 
located about $\vec{r}=0$, at $\pm \vec{r_1}$. An  anti-D3-brane is 
located at the bottom of each 
throat. The D3-brane is located in the vicinity of the point of $Z_2$ symmetry, 
at $\vec{r}=0$. See Figure 1. 
\begin{figure}
\begin{center}
\epsfig{file=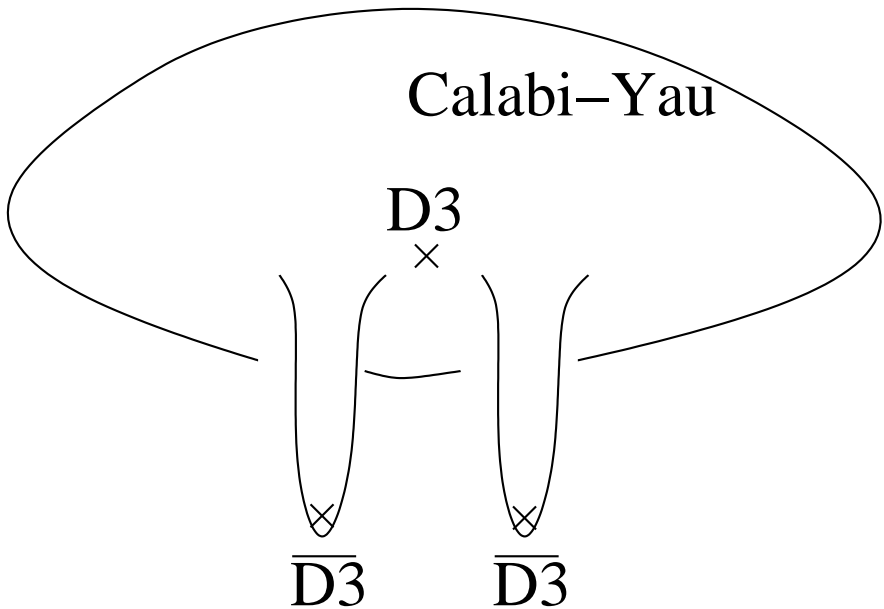,height=80mm}
\end{center}
\caption{Two symmetrically located Klebanov-Strassler throats in Calabi-Yau space. 
Anti-D3-branes are at the bottom of each throats and mobile D3-brane is in between.}
\end{figure}
We are now ready to include the effects of the Coulomb interaction in 
eq.~(\ref{babpot}) in such a set-up\footnote{Strictly speaking the $Z_2$ symmetry is not essential for this model. 
The idea can work for any two appropriately positioned throats, as will 
be discussed further in section 3, when we consider generalizations.}.
For simplicity we assume that the metric $g_{mn}$ is flat and this is consistent with the
 form of the K\"ahler potential assumed above in eq.~(\ref{exppot}).

The second term in eq.~(\ref{babpot}) is then given by 
\be
\label{intpot}
V^I=-2T_3^2 Z^8 {1 \over 2\pi^3 M_{10}^8} ({1 \over |\vec{r}-\vec{r_1}|^4} + 
{1 \over |\vec{r}+\vec{r_1}|^4} ) \,.
\ee
$Z^4=e^{4A}$ is the redshift factor at the location of either anti-brane,
and by symmetry this is the same.
Expanding to quadratic order, and using the relation, ${T_3^2\over M_{10}^8}=\pi$ we get, 
\be
\label{eintpot}
V^I=-{2Z^8 \over \pi^2 r_1^4} \bigl(1 +{2 \over r_1^2}
({6r_{1m}r_{1n} \over r_1^2} -\delta_{mn}) r_m r_n \bigr) \,.
\ee
The first term on r.h.s. gives a correction to the vacuum energy. 
We will see below that this correction is small. 
The second term  is quadratic in the displacement, $\vec{r}$,
and gives a contribution to the mass. If the anti-branes  
are located  along the $y^1$ direction, it takes the form, 
\be
\label{quadpot}
V^I_{quad}=-{4 Z^8 \over \pi^2 r_1^6} \bigl(5(y^1)^2 - (y^2)^2 -(y^3)^2
-(y^4)^2-(y^5)^2 -(y^6)^2 \bigr) \,.
\ee
In particular the mass term associated with the $y^1$ direction is negative, 
as would be expected from the attractive nature of the force. 

The kinetic energy terms for brane moduli can be derived from the DBI action,
\be
\label{DBI}
L=-T_3\int d^4x \sqrt{-g} {1 \over 2} g^{\mu\nu}\partial_\mu y^i \partial_\nu y^i \,.
\ee
From this we see that the  canonically normalized field
$\varphi^i=\sqrt{T_3}y^i$.

Requiring the negative mass term along the $y^1$ direction to approximately cancel 
the curvature induced term discussed in the previous section then gives the condition, 
\be
\label{conda}
{2\over 3} {V_0(\rho_c) \over M_{Pl}^2}  \simeq {40 Z^8 \over \pi^2 r_1^6} {1 \over T_3} \,.
\ee

As was mentioned above, up to a factor of unity, the de Sitter vacuum energy is of order
the contribution of the anti-branes, $V_0(\rho_c) \sim T_3 Z^4$. Dropping factors of order unit we then get,
\be
\label{condb}
{T_3^2 Z^4 \over M_{Pl}^2} \sim { Z^8 \over r_1^6} \,.
\ee
Using the relations, $M_{Pl}^2=M_{10}^8 L^6$, and $T_3^2/M_{10}^8=\pi$, this leads finally to 
the condition, 
\be
\label{condf}
r_1 \sim  Z^{2/3} L \,. 
\ee
Note that since $Z < 1$,  eq.~(\ref{condf}) is  consistent  with 
requiring  that  the brane-anti-brane separation is less than size of the compactification, $r_1<L$.

Let us make two comments before closing this subsection. 
First, as was mentioned above, $Z$ in eq.~(\ref{condf}) scales with the volume.
E.g., for a K-S throat, $Z^4 =e^{4A} \sim {L^4 \over (\alpha^{'})^2} \exp(-{8 \pi K \over 3 g_s M})$,
where, $M$ and $K$ are two integers which specify the flux of $F_{3}$ and $H_{3}$ threading the 
vanishing $S^3$ and its dual three-cycle respectively \cite{GKP,KS}.  
Including this in eq.~(\ref{condf}) and requiring $r_1 <L$, gives rise to the condition, 
\be
\label{condfalt}
e^{-\left({4\pi K \over 9 g_s M}\right)} < \left({\sqrt{\alpha^{'}} \over L} \right)^{2/3} \,,
\ee
which can be met for $L>\sqrt{\alpha^{'}}$ by choosing appropriate integers $K,M$. 
Thus we see that by appropriately choosing the fluxes, the two contributions to the mass, 
from the curvature coupling and the brane-anti-brane interaction, can be made to approximately cancel. 
Second, we mentioned above that the first term in eq.~(\ref{eintpot}) makes a small contribution to the vacuum energy. 
We can now verify this. We had mentioned above  that the vacuum energy is of order the anti-brane potential 
energy, i.e., $V_0(\rho_c) \sim T_3 Z^4$. So  the required condition is, $T_3 Z^4 \gg {Z^8 \over r_1^4}$.
  From eq.~(\ref{condf}) this takes the form,  ${1\over g_s} {L^4\over (\alpha^{'})^2} \gg Z^{4/3}$,
where we have used the relation, $T_3 \sim {1\over g_s (\alpha^{'})^2}$. This is obviously met when, 
 $g_s < 1 $,   ${L \over \sqrt{\alpha^{'}}} >1$ and $Z \ll 1$. 

In the following discussion we will assume that the fluxes etc have been chosen so 
that  eq.~(\ref{condf}) is met and 
 the two contributions to the mass for the inflaton approximately
 cancel, leaving a small  residual negative $({\rm mass})^2$,  for motion along the $y^1$ direction. 
The point $\vec{r}=0$, is then a maximum of the potential along this direction.
In its vicinity the potential is  given by, 
\be
\label{poti}
V=V_0 -{1 \over 2} m^2 \phi^2 \,,
\ee
where $V_0\equiv V_0(\rho_c)$, $\phi \equiv \sqrt{T_3}y^1$ is the canonically normalized field,
and $m^2$ is the small residual mass.
We will examine the inflationary scenario that results when one starts close to this maximum next. 
Along all the other directions the potential is stable. For simplicity we will assume that 
during inflation the 
brane is at rest  along these directions, at
$y^i=0, i=2, \cdots, 6$.
Before proceeding let us note that inflation with a potential of the form eq.~(\ref{poti})
was studied earlier in \cite{German:2001tz}. 
This paper noted that the scale of inflation would have to be low 
and also the tilt would be negative in accord with our analysis in the next subsection.

%%%%%%%%%%%%%%%%%%%%%%%%%%%%%%%%%%%%%%%%%%%%%%%%%%%%%%%%%%%%%%%%%%%%%%%%%%%%%%%%%%%%%%%%%%%%%%%%%%%%
\subsection{The Resulting Inflationary Scenario}
%%%%%%%%%%%%%%%%%%%%%%%%%%%%%%%%%%%%%%%%%%%%%%%%%%%%%%%%%%%%%%%%%%%%%%%%%%%%%%%%%%%%%%%%%%%%%%%%%%%%

The potential for the inflaton was described in the previous section, eq.~(\ref{poti}). 
In the standard classification of inflationary models this is a  canonical example of ``small field"
inflation (in the classification of \cite{DK}, see also \cite{KK}).

The Hubble scale during inflation is,  
\be
\label{valh}
H^2={V \over 3M_{Pl}^2} \simeq {V_0 \over 3M_{Pl}^2} \,.
\ee
The last  approximation follows from the fact that $\phi^2 \ll {V_0^2 \over m^2}$, during inflation,
as we will see shortly. 
The two slow roll parameters are given by 
\be
\label{vale}
\epsilon \equiv {1 \over 2} \left({V^{'} \over V}\right)^2 M_{Pl}^2 ={1\over 18}\left({m^2\over H^2}\right)^2 
\left({\phi^2 \over M_{Pl}^2}\right) \,,
\ee
and
\be
\label{valeta}
\eta \equiv {V^{''} \over V} M_{Pl}^2 = {m^2 \over 3 H^2} \,.
\ee
The condition for slow-roll inflation is that  $\epsilon, \eta \ll 1$.
We see that by taking
\be
\label{condmass}
m^2 \ll H^2 
\ee
and starting at a small enough value of $\phi$, both
these conditions can be met.
 
Let us examine the observational constraints imposed in more detail now. We will see that they 
determine the scale of inflation,
independent of the  string coupling constant 
and the size of Calabi-Yau space in this model.   

The slow roll conditions stop holding
 in our model when the subleading corrections to the potential become 
significant. We denote  the value of the inflaton when this happens by, $\phi_f$.
 There are two sources for these corrections.  From eq.~(\ref{intpot}), 
we see that there are corrections\footnote{Here we are using the fact that the canonically normalized field 
$\phi=\sqrt{T_3} r$.}
of ${\cal{O}}({\phi \over\sqrt{T_3} r_1})^2$. These will become important when 
\be
\label{end}
\phi \sim \sqrt{T_3} r_1 \,.
\ee
From eq.~(\ref{exppot}),  we see that there are also corrections of 
${\cal{O}}({\phi \over M_{Pl}})^2$, these become significant when $\phi \sim M_{Pl}$.
It is easy to see that eq.~(\ref{end}) is more restrictive. This follows from 
 eq.~(\ref{condf}) after noting that,  $\sqrt{T_3} r_1 
\sim \sqrt{T_3} Z^{2/3} L \ll M_{Pl}=M_{10}^4 L^3$, for $g_s < 1$, $Z\ll 1$ and $L/\sqrt{\alpha^{'}} >1$.
So we learn that inflation comes to an end when $\phi_f\sim \sqrt{T_3}r_1$. 
This condition of course makes good physical sense.
The anti-brane is located at $\phi=\sqrt{T_3} r_1$, and  we  expect inflation to have ended    
by the time the brane gets to the vicinity of the anti-brane. 

Two comments are worth making at this stage. 
The two  conditions,  $\phi \le \phi_f \ll M_{Pl}$,
and, eq.~(\ref{condmass}),
imply that   
$\phi^2 \ll {V_0 \over m^2} \simeq {H^2 \over m^2} M_{Pl}^2$, as was mentioned above eq.~(\ref{vale}).
Also, using  these two conditions  in   
eq.~(\ref{vale}) and  eq.~(\ref{valeta}), we learn that  
 $\epsilon \ll \eta$. The observed value of the tilt, as seen below in eq.~(\ref{tilt}),
then tells us that 
\be
\label{valmass}
\eta ={m^2 \over 3 H^2} \sim - 10^{-2} < 0 \,.
\ee 

It is easy to see that the value of the inflaton, $\phi$, 
 $N_e$ $e$-foldings before the end of inflation, is given by
\be
\label{relp}
\log\left({\phi_f \over \phi}\right) ={|m|^2 \over 3 H^2} N_e \,.
\ee
The scale of the adiabatic density perturbations, \cite{Liddle}, \cite{LythRiotto}, is given by
\be
\label{sdp}
\delta_H={1 \over \sqrt{75} \pi}{1 \over M_{Pl}^3}{V^{3/2}\over |V^{'}|} = {3 \over 5 \pi }{H^3 \over |m^2 \phi|} \,.
\ee
Data tells us that $\delta_H = 1.9 \times 10^{-5}$, see for example, \cite{WMAP}.
The tilt in the spectrum is given by
\be
\label{tilt}
n = 1 - 6 \epsilon + 2 \eta \simeq 1 + 2 \eta \,.
\ee
 From eq.~(\ref{valmass}), we see that this is less than unity  in our model. 
Observational data indicates a non-zero tilt, with, $n \simeq 0.97$, \cite{WMAP}, \cite{WMAP2}.

We are now ready to consider the constraint imposed by the density perturbations, eq.~(\ref{sdp}).
The observed anisotropy arises due to perturbations which leave the horizon about $60$ $e$-foldings before the
end of inflation\footnote{The  number of $e$-foldings depends on the reheat temperature.
While we leave this matter for further study, it seems likely that the reheat temperature 
in this model will turn out to be low, since  the Hubble scale is quite low,  $H =10^{10}$ GeV. 
This decreases $N_e$, but does not change the estimates below significantly.}. 
From eq.~(\ref{relp}), eq.~(\ref{valmass}), we learn that the value of the inflaton, 
$60$ $e$-foldings before the end of inflation, is given by,
$\phi \sim \phi_f/1.8 \sim \sqrt{T_3}r_1/1.8$.
Substituting this\footnote{Precisely speaking, this $N_e$ $e$-foldings suppression factor $1.8 \simeq \exp(N_e |\eta|)$ 
is dependent on $\eta$, but this $\eta$-dependence is very weak 
so we will neglect this dependence in the following discussion.} 
along with eq.~(\ref{valmass}), in eq.~(\ref{sdp}), then gives us,
\be
\label{condh}
{H \over \sqrt{T_3} r_1} \sim {1 \over 1.8} \times 10^{-4}{|m|^2 \over H^2} 
\sim {3 \over 1.8} \times 10^{-4} \hspace{1mm} |\eta| \,.
\ee
Some more algebraic manipulation leads to a determination of the Hubble scale during inflation. 
Using the relation, $V_0(\rho_c) \sim T_3 Z^4$,  and eq.~(\ref{valh}), one gets, 
\be
\label{condaa}
{1 \over \sqrt{3}}{Z^2 \over r_1 M_{Pl}} \sim {3 \over 1.8} \times 10^{-4} \hspace{1mm} |\eta| \,.
\ee
Next substituting eq.~(\ref{condf}), in eq.~(\ref{condaa}),  leads to the relation
\be
\label{condab}
{1 \over \sqrt{3}}{Z^{4/3} \over L M_{Pl}} \sim {3 \over 1.8} \times 10^{-4} \hspace{1mm} |\eta| 
\ee
which determines the redshift factor, $Z$, in terms of the scale of compactification $L$ and $M_{Pl}$.
Finally, putting this condition into the expression for the Hubble constant yields, 
\be
\label{condfh}
{H^2 \over M_{Pl}^2} \sim {T_3 Z^4 \over 3 M_{Pl}^4} \sim 0.8 \times 10^{-11} {T_3 \over M_{10}^4} |\eta|^3 
\sim 1.4 \times 10^{-17} \left({|\eta| \over 0.01}\right)^3 \,,
\ee
where we have used the relation, $M_{Pl}^2 = M_{10}^8 L^6$, and ${T_3 \over M_{10}^4} = \sqrt{\pi}$.
Note that various model dependent features like the scale of the compactification of Calabi-Yau space and 
the value of $g_s$ drop out in this expression. 
The resulting value of the Hubble scale is indeed low in this model.
Eq.~(\ref{condfh}) gives,
\be
\label{valht}
H \sim 9.2 \times 10^{9} \left({|\eta| \over 0.01}\right)^{3/2} \mbox{GeV} \,.
\ee 
This corresponds to an energy scale  
\be
\label{vallambda}
\Lambda \sim 2.0 \times 10^{14} \left({|\eta| \over 0.01}\right)^{3/4} \mbox{GeV} \,,
\ee
which is small compared with 
the  SUSY GUT scale,  $ 10^{16}$ GeV.

The power in gravity wave perturbations is given by, 
\be
{\cal{P}}_{grav} = {1 \over 2 \pi^2} {H^2 \over M_{pl}^2} \sim 7 \times 10^{-19} \left({|\eta| \over 0.01}\right)^3 \,. 
\ee
It is clear that the production of gravity waves in this model is greatly suppressed,
much below the level of detection in future experiments.

To summarize, this model gives rise to an example of small field inflation\footnote{ This is an ``A" type model
in the classification of \cite{WMAP2}.}.
The inflaton varies by much less than the Planck scale during inflation. 
The slow roll parameter $\epsilon$ is extremely small and the tilt is determined by $\eta$ which is negative. 
The Hubble scale during inflation is quite low, of order $10^{10}$ GeV, 
and the corresponding vacuum energy is of order $10^{14}$ GeV.
As a result, the observed anisotropy arises almost entirely due to adiabatic density perturbations and  
the production of gravity waves is highly suppressed.

Let us end this section with some  comments.
The qualitative features of the inflationary scenario in this model mainly arise from the broad maximum,
with $m^2/H^2 \ll 1$,  and are  quite insensitive to various details and approximations.
For example, we assumed at various points in the discussion that the internal metric, eq.~(\ref{warpmet}), is
trivial, $g_{mn}=\delta_{mn}$. This is of course an approximation valid only locally, since Calabi-Yau spaces 
are not flat.  For a non-trivial metric the $Z_2$ symmetric point will still be an extremum. 
 It will always be a maximum of the attractive  Coulomb potential  eq.~(\ref{intpot}).
In addition if it is a minimum for the terms, eq.~(\ref{potb}),
the basic idea will work. 
By adjusting the redshift factor, $Z$, as a function of
the brane-anti-brane separation, as in eq.~(\ref{condf}),
one can  arrange  a near cancellation, leading to the condition
$m^2/H^2 \ll 1$.
It could be that the corrections to the metric causes inflation to end for a smaller value of $\phi$ than we
estimated above. It follows from eq.~(\ref{condh}),
that the resulting Hubble scale of inflation will  be lower than our
estimate above, order $10^{10}$ GeV.

It is interesting to compare the inflationary parameters obtained above, with those obtained in 
the KKLMMT model, with the curvature induced mass term set to zero by hand, 
appendix C, \cite{KKLMMT}. In the  latter  case after  setting $N_e=60$, one finds that $\delta_H$
directly determines the energy scale during inflation, $\Lambda$, (C.12), \cite{KKLMMT}. 
Remarkably, the resulting value 
is  the same as that obtained above, $10^{14}$ GeV. 
The tilt parameter is directly determined by $N_e$, and with $N_e=60$, is
  in good agreement with the 
data, (C.19), \cite{KKLMMT}. In contrast in the model above, 
we saw that the energy  scale during inflation and the tilt are  relatively insensitive to $N_e$. 
Fixing  the the value of $m^2/H^2$, so that the tilt agrees with observations, we found  that  
$\delta_H$  determines  the energy scale during inflation.

\subsection{Parameter Constraints}

{\it 2.5.1. Meeting the Various Constraints:}

There are four important parameters which characterize the compactification, 
$r_1$ - which governs the distance between the
two K-S throats, $Z$, the redshift at the bottom of each K-S throat,  $L$, the size of the compactification,
and $g_s$ the string coupling. Here we  will examine the various constraints imposed on them.
As we will see  these turn out to be very  stringent and will  restrict some of the parameters  
 to within an order of magnitude or so. Our estimates are uncertain by factors of order unity because of 
 our lack of knowledge about how to define some of the scales  precisely.
 Within these uncertainties, we will find  that all 
the constraints can be met. This establishes, as a first pass, that this  model is viable. 
However, given the stringent nature of the constraints one would like to do better. 
This requires an improved  estimate of the numerical factors in the constraints and  is not easy.
For some of the constraints one will probably need explicit string theory constructions of the model.
We leave this for the future.

The constraints on the parameters arise in three ways. First, the  low-energy supergravity theory,
 within which our analysis has been carried out, must be valid. Second,  the initial brane-anti-brane 
separation, $r_1$,  must satisfy some conditions so that the  form of the potential 
we assumed in eq.~(\ref{intpot}) is valid and the resulting maximum is broad.  
Third, the observational constraints discussed in section 2.4 must be met. 
We take these up in turn now. 

For the  low-energy supergravity  approximation to be valid,  the $\alpha^{'}$ and $g_s$, expansions must hold. 
These give rise to the conditions
\be
\label{al} 
L^6 \gg  (\alpha^{'})^3 \,,
\ee 
and
\be
\label{qu}
g_s \ll 1 \,,
\ee
respectively. 
 In addition,  the scale of supersymmetry breaking must be  small
compared to the string scale. 
We take this condition to be 
\be
\label{suba}
T_3 Z^4 \ll  {1 \over (2\pi)^3 (\alpha^{'})^2} \,.
\ee
The l.h.s. above arises because the scale of supersymmetry breaking is set by the anti-brane tension.
The r.h.s. has been chosen as follows. We expect the supersymmetry breaking scale to be unacceptably 
large when $g_s=Z=1$. The r.h.s. is the value of the D3-brane tension for this choice\footnote{The 
numerical factor, ${1 \over (2\pi)^3}$, is less than unity and therefore makes eq.~(\ref{suba}) more stringent.} of $g_s,Z$. 
Eq.~(\ref{sub}) can be then  re-expressed as a condition on $Z,g_s$ alone, and takes the form 
\be
\label{sub}
Z^4 \ll g_s \,.
\ee

The initial brane-anti-brane separation $r_1$ must satisfy three conditions.
First, it must be big enough so that no tachyon is present at the start of inflation. 
The mass of the tachyon $m_{T}^2$ as a function of the initial separation, $r_1$, is given by
\be
\label{mass}
m_{T}^2={r_1^2 \over (2\pi \alpha^{'})^2}-{1\over 2 \alpha^{'}} \,.
\ee
The requirement that no tachyon is present then   takes the form, 
\be
\label{nt}
{r_1 \over \sqrt{\alpha^{'}}} \gg \sqrt{2} \pi \,.
\ee
Second,  $r_1$ must be small enough   compared to the size of the  compactification, $L$, so 
 that the
corrections to the harmonic function due to compact nature of the internal space can be neglected. 
Third, the resulting maximum should be broad, this gives eq.~(\ref{condf}). 
In the analysis below we will assume that eq.~(\ref{condf}),
 is the more restrictive of the latter  two constraints. 
In the particular examples we consider we will see this is true.  

The observed anisotropy gives the  condition, eq.~(\ref{condab}).
Also, the tilt, eq.~(\ref{tilt}),  determines $\eta$ and thereby $m^2/H^2$ by eq.~(\ref{valmass}).
It is worth repeating here, that eq.~(\ref{condab}) completely fixes the energy scale during inflation 
as eq.~(\ref{vallambda}), 
or equivalently, the SUSY breaking scale up to a factor of order unity. 

To summarize  the four parameters, $r_1, Z, L, g_s$, mentioned above,   must meet the constraints, 
 eq.~(\ref{al}), eq.~(\ref{qu}), eq.~(\ref{sub}),
eq.~(\ref{nt}), eq.~(\ref{condf}), and, eq.~(\ref{condab}).

We are now ready to analyze these constraints in more detail.

{\it 2.5.2. Analysis of the Constraints:}

We can view, eq.~(\ref{condab}) as determining $Z$ in terms of $L$, and then eq.~(\ref{condf})
as determining $r_1$ in terms of $L$.
This leaves $L$ and $g_s$ undetermined. They must  satisfy the remaining four constraints, 
 eq.~(\ref{al}), eq.~(\ref{qu}), eq.~(\ref{sub}), and eq.~(\ref{nt}).

Since
\be
\label{valmten}
M_{Pl}^2=M_{10}^8 L^6 \,,
\ee
 eq.~(\ref{condab}) takes the form
\be
\label{condtemp}
Z^4 \simeq {3^{9/2} \over (1.8)^3} \times 10^{-12} (M_{10} L)^{12} |\eta|^3 \,.
\ee
 Eq.~(\ref{sub}) then imposes the condition, 
\be
\label{consa}
(LM_{10})^{12} \ll {(1.8)^3 \over 3^{9/2}} \times 10^{12}{g_s \over |\eta|^3} \,.
\ee
Using the relation 
\be
\label{valls}
{1\over M_{10}^8}= {1\over 2} (2\pi)^7 (\alpha^{'})^4 g_s^2 \,,
\ee
 eq.~(\ref{al}) takes the form, 
\be
\label{const}
(LM_{10})^{12} \gg {1\over (2^6 \pi^7)^{3/2}} {1\over g_s^3} \,.
\ee
Similarly, using eq.~(\ref{valls}) and eq.~(\ref{condf}),
 eq.~(\ref{nt}) gives
\be
\label{consb}
(LM_{10})^{12} \gg {(1.8)^2 \sqrt{\pi} \over 3^3 \cdot 2} {10^8\over |\eta|^2 g_s} \,.
\ee
It is easy to see that for reasonable values of $g_s$, eq.~(\ref{consb}) is more restrictive than 
 eq.~(\ref{const}). 

So we see that $L$, the size of the compactification, and $g_s$, the string coupling, 
 must satisfy the  conditions, eq.~(\ref{consa}), eq.~(\ref{consb}), and eq.~(\ref{qu}).

We can express eq.~(\ref{consa}) and eq.~(\ref{consb}) together as,
\be
\label{finalcond}
1 \times 10^{12} \Bigl({g_s\over 0.1}\Bigr)^{-1} \Bigl({|\eta| \over 0.01}\Bigr)^{-2} \ll (M_{10} L)^{12} \ll 
 4 \times 10^{15}
\Bigl({g_s \over 0.1}\Bigr)\Bigl({|\eta| \over 0.01}\Bigr)^{-3} \,.
\ee
Since the upper and lower bounds in the above inequality are somewhat far apart,  
for reasonable values of $g_s$ and $\eta$, we
 see that the required constraints on $L$ can be met.

To summarize, we saw in the analysis above that $r_1$ and $Z$ can be expressed in terms of $L$ and $|\eta|$, 
using the relations eq.~(\ref{condf}) and eq.~(\ref{condab}). 
The remaining two parameters, $g_s$ and $L$ must then satisfy the condition 
eq.~(\ref{qu}) and  eq.~(\ref{finalcond}). We saw above that  these can be met.

{\it 2.5.3. Explicit Examples with Conclusions:}

It is worth examining some  explicit examples which meets all the constraints in more detail.

{\it Scenario I}

We take $(M_{10}L)^{12}=10^{13}$, with  $g_s=0.1$ and  $|\eta|=0.01$. Note that 
$(M_{10}L)^{12}=10^{13}$ lies between the two bounds in eq.~(\ref{finalcond}). 
Using, eq.~(\ref{valmten}), we now obtain, $M_{10}=1.4 \times 10^{15}$ GeV,
$1/L = 1.1 \times 10^{14}$ GeV. The redshift factor is then given by  eq.~(\ref{condtemp}). For $|\eta| = 0.01$ we get, 
$Z^4=2.4 \times 10^{-4}, Z = 0.12$. The brane-anti-brane separation, $r_1$ given by eq.~(\ref{condf}), 
to be $1/r_1=4.5 \times 10^{14}$ GeV. Using 
eq.~(\ref{valls}), with $\alpha^{'}=l_s^2$, we get the string scale as $1/l_s=3.5 \times 10^{15}$ GeV.
Finally, as we mentioned above, the SUSY breaking scale, $(T_3 Z^4)^{1/4}$, 
 is of order the vacuum energy during inflation, $2.0 \times 10^{14}$ GeV. 
These different energy scales are summarized as follows:
\begin{center}
{\bf Table I \hspace{10mm}}
\begin{tabular}{cr}
physical quantities & Scale (GeV) \\
$M_{pl}$          & $2.4 \times 10^{18}$ \\
$1/l_s$           & $3.5 \times 10^{15}$ \\
$M_{10}$          & $1.4 \times 10^{15}$ \\
$1/r_1$           & $4.5 \times 10^{14}$ \\
$\Lambda=(T_3 Z^4)^{1/4} \hspace{5mm}$ & $2.0 \times 10^{14}$ \\
$1/L$             & $1.1 \times 10^{14}$ \\
$H$               & $0.9 \times 10^{10}$ \\
\end{tabular}

warped factor: $Z = 0.12$ 

 $g_s = 0.1$, $|\eta| = 0.01$
\end{center}

{\it Scenario II}

To get an idea of how these parameters change, we also consider the another scenario
where $(M_{10}L)^{12}=10^{14}$ and $g_s=0.1$. For $|\eta|=0.01$, the resulting values for the parameters 
are summarized in Table II. One sees that qualitatively the energy scales look similar, although some of 
them become a little bit smaller.

\begin{center}
{\bf Table II \hspace{10mm}}
\begin{tabular}{cr}
physical quantities & Scale (GeV) \\
$M_{pl}$          & $2.4 \times 10^{18}$ \\
$1/l_s$           & $2.0 \times 10^{15}$ \\
$M_{10}$          & $7.7 \times 10^{14}$ \\
$1/r_1$           & $1.4 \times 10^{14}$ \\
$\Lambda=(T_3 Z^4)^{1/4} \hspace{5mm}$ & $2.0 \times 10^{14}$ \\
$1/L$             & $5.2 \times 10^{13}$ \\
$H$               & $0.9 \times 10^{10}$ \\
\end{tabular}

warped factor: $Z = 0.22$ 

 $g_s = 0.1$, $|\eta| = 0.01$
\end{center}

Finally we note that some limited amount of variation is allowed for $g_s$. In general reducing
$g_s$ tightens the constraints in eq.~(\ref{finalcond}), increasing the lower bound and decreasing the upper bound.
E.g., taking $g_s=0.01$, the lower bound in eq.~(\ref{finalcond}) becomes $1 \times 10^{13}$,
and the upper bound be $4 \times 10^{14}$. 

From Table I and II, we see that the parameters are quite tightly restricted by the constraints. 
While all the inequalities we discussed above are indeed met, several relevant scales are close together. 
Thus, our analysis does not conclusively establish that  the required conditions have been met.
For example in Table II, $1/r_1=1.4 \times 10^{14}$ GeV, while $1/l_s=2.0 \times 10^{15}$ GeV, so that  
 $r_1 /(\sqrt{2} \pi \sqrt{\alpha^{'}}) = 3.1 > 1$, as is needed for the absence of a tachyon. 
 However, since our estimate of $r_1$ in eq.~(\ref{condf}) is uncertain by a factor of order unity,
 and the above ratio is not much bigger than unity, a more careful estimate is needed to conclusively establish this point.

As we mentioned at the beginning of this subsection, a more careful study is therefore
needed to  establish that the constraints are satisfied.  This would include a better 
understanding of the requirements for the low-energy supergravity approximation to be valid.
It would also need to be done in the context of concrete constructions in string theory.
In such constructions one can hope to calculate the numerical factors in eq.~(\ref{condf}) and eq.~(\ref{condab}), 
and also understand whether the brane-anti-brane potential is well approximated by eq.~(\ref{intpot}). 
Finally,  a better  observational  determination
of the tilt parameter, and therefore of  $\eta$, will also help. 

Let us end with some comments. 
From Table II we see that $r_1/L\simeq 1/(2.7)$. Corrections to the potential, eq.~(\ref{intpot}),
 due to compact nature of the internal
space can be estimated roughly as arising due to ``images". These contributions 
are small and do not affect eq.~(\ref{condb}) 
since the harmonic function falls like $1/r^4$ in six dimensions. 

We see from Table I and II that  $1/L \gg H$, so that the internal space has a size much 
bigger than the Hubble scale. This is consistent with our four-dimensional description of inflation.

It is worth pointing out that the string scale in our model is close to  the SUSY GUT scale but slightly lower. 
This connection is interesting to explore further. It will be interesting if successful inflation
necessarily requires a string tension of order the SUSY GUT scale. 

We have assumed above that the parameters, $Z$, $r_1$, $L$, $g_s$, can in effect be tunned independently
to meet all the constraints. This needs to be checked in explicit models. 
Here we only note that given the large number of fluxes which we can turn on and dial, 
this seems plausible. 

The important observational constraint in the analysis above  comes from the scale of density perturbations 
eq.~(\ref{condab}).  Requiring a small scale of density perturbations make the r.h.s. of
eq.~(\ref{condab}) smaller, and   the upper bound in eq.~(\ref{finalcond})  bigger.
As a result the constraint eq.~(\ref{finalcond}) becomes less severe\footnote{One way to make it
 easier to get 
the   required density perturbations  would be to end inflation earlier. Since 
$\delta_H \sim H^3/(m^2\phi)$,
smaller $\phi$ gives  bigger $\delta_H$. This can  be done if the moving D3-brane encounters a 
D7-brane
or some other feature which obstructs its motion  before it gets close to the anti-D3-brane.}. 
It therefore seems likely that even after a more detailed analysis along the lines mentioned above,
our model will remain  viable as an example of slow-roll inflation in string theory.
By varying the different fluxes etc one can probably implement the model in many different ways in
string theory, giving rise to inflationary scenarios with different values for the density perturbations,
tilt etc. Hopefully, some of these will agree with the data, although the vast majority will probably not.

%%%%%%%%%%%%%%%%%%%%%%%%%%%%%%%%%%%%%%%%%%%%%%%%%%%%%%%%%%%%%%%%%%%%%%%%%%%%%%%%%%%%%%%%%%%%%%%%%%%%
\section{Discussion}
%%%%%%%%%%%%%%%%%%%%%%%%%%%%%%%%%%%%%%%%%%%%%%%%%%%%%%%%%%%%%%%%%%%%%%%%%%%%%%%%%%%%%%%%%%%%%%%%%%%%

{\it 1. Putting the Proposal on Firmer Footing:}

Some important issues need to be studied further to put this model on a firmer footing. 
Two of these are common to many  KKLT type constructions.  First,  
non-perturbative effects, like gaugino condensation,  responsible for volume stabilization need to be 
understood better.
Second, in deriving the full potential we  assumed that the brane-anti-brane potential can be simply 
added to the contribution coming
from the superpotential. This would be true if the former arises through a $D$-term. For evidence in support 
of this see \cite{Dvali:2003zh,Binetruy:2004hh}. Finally, 
as mentioned above, explicit examples of Calabi-Yau spaces need to be constructed, where all the
constraints discussed above can be met simultaneously. 
For all these reasons the construction discussed in this paper should be thought of as a proposal rather than a
concrete model at the moment. 

For recent progress towards meeting some of the requirements of the KKLT construction see, 
\cite{GKTT}. 

{\it 2. Building a More Complete Model:}

As a model of inflation our construction is incomplete in three important ways. 
 We have not addressed how inflation ends, and how it begins. And we have not asked if the standard model 
can be satisfactorily incorporated in this construction.
Let us briefly comment on some of these issues here.

{\it 2a)  Ending Inflation:}
This model is in fact an example of hybrid inflation. 
When the brane-anti-brane separation gets to be of order the string scale, a tachyon develops.
What happens next is currently a matter of active study. See for example 
\cite{Sen:2002nu,Cline:2002it,Lambert:2003zr,Gaiotto:2003rm,Sen:2003xs,Sen:2003mv}. 
 It seems reasonable to believe that the 
energy in the brane-anti-brane pair is eventually converted to light closed string modes like the graviton. 
Whether this energy can be efficiently transferred to the standard model degrees of freedom,
reheating the universe satisfactorily, is   dependent on how the standard model 
couples to the degrees of freedom  in the inflationary  throat. This is tied to another
incomplete aspect of the model mentioned above, namely how the standard model is incorporated in it, and 
needs to be studied further.

Cosmic strings, both of D and F type will be produced in this model at the end of inflation
due to D3 anti-D3-brane annihilation \cite{Sen:1999mg}, \cite{Tyestrings}. If these strings are metastable
\cite{Copeland:2003bj}, their tension must meet the condition, 
$G_N T_1 \le 10^{-7}$, to avoid generating unacceptably large  anisotropies. 
A D1-brane at the bottom of the K-S throat 
has tension,  $T_{D1}={1\over 2\pi g_s \alpha^{'}} Z^2$, where the $Z$ dependence arises due to the redshift.
The tension of the anti-D3 brane is given by $T_3 Z^4$ and is of order the vacuum energy during inflation 
$(2 \times 10^{14} \hspace{1mm} {\rm GeV})^4$.  
Using the relation $T_{D1}=\sqrt{2\pi T_3 Z^4 / g_s}$, $T_{F1}=\sqrt{2\pi T_3 Z^4  g_s}$,
 we get 
$G_N \sqrt{T_{D1} T_{F1}} \simeq  6.5 \times 10^{-10}$. 
This is lower than the bound mentioned above. Interestingly,  future observations
might be sensitive to  such low values of the tension, \cite{Copeland:2003bj}, \cite{gwaves}.

A small positive cosmological constant is needed after the end of inflation to account for the acceleration
of the present epoch. This could arise as follows. The brane is drawn towards one of the throats 
and annihilates the anti-brane at the bottom of it. But the anti-brane in the second throat survives after inflation ends.
The final resulting cosmological constant will get a contribution 
from this anti-brane's tension and the superpotential terms, eq.~(\ref{potenr}). These 
can  nearly  cancel leaving behind 
a small positive value.

{\it 2b) The Initial Conditions and Eternal Inflation}

The scale of inflation in this model, $\Lambda \sim 10^{14}$ GeV, is considerably smaller than
the ten-dimensional Planck scale, and   makes the question of initial conditions all the more important.

We can only offer a few speculations about this here. It is possible  that the universe did not 
begin in  a big bang, with temperatures of order the Hagedorn temperature.  Instead as has been suggested 
in \cite{Vilenkin,lindeeter},
it might have begun by a tunneling event in the vicinity of the maximum of the potential. 

It could also be that the question of initial conditions is not very significant in this model. 
It has been argued that  a potential with a broad maximum, of the kind we have here, will
give rise to eternal inflation. 
If the inflaton is close enough to the top of the potential of eq.~(\ref{poti}) with $\phi < \phi_c \sim H^3/m^2 $,
quantum fluctuations can  drive it up the hill at a faster rate than the classical
gradient term allows it to descend. 
Since the regions where the cosmological constant is bigger also grow exponentially more rapidly, 
soon the universe will be dominated by regions where the inflaton is at the top of the hill,
making the initial conditions irrelevant.
The observed universe in this picture would arise when 
fluctuations cause the inflaton  to descend far enough from the top so that the classical evolution
discussed in section 2.4, becomes valid. 
This is an attractive and fairly plausible picture, but it is somewhat speculative at the moment and 
needs to be understood better.

{\it 3. Generalizations of the Model:}

{\it 3a) Asymmetric Throats:}

As was mentioned towards the end of section 2.4, the inflationary scenario implemented in this note is
mainly dependent on the small curvature at the maximum and independent of many details of the model.
For example, we have considered the $Z_2$ symmetric model in our discussion above, but the
$Z_2$ symmetry is not essential.
In general, the curvature couplings discussed in section 2.2 will result in a potential for the inflaton.
Suppose this potential has a minimum at some point in the compactification.  
Two K-S-like throats containing anti-branes at the bottom, not necessarily symmetrically located, whose 
distance from the minimum is adjusted relative to the red-shift factors  and the number of anti-branes 
at the bottom of the throats, can then  
nearly cancel the resulting mass giving rise to a maximum with a small curvature. 
The resulting inflationary scenario will then be qualitatively unchanged, in particular the scale of inflation
will continue to be quite low.

{\it 3b) More Symmetry:}

Conversely we can have situations with higher symmetry, with more than two throats symmetrically located with respect
 to a point of enhanced symmetry. The term $({6r_{1m} r_{1n} \over r_1^2} -\delta_{mn})$ in eq.~(\ref{eintpot}),
would then be replaced by
the quadrapole moment of the resulting configuration, $Q_{mn}$. As long as $Q_{mn}$ does not vanish, it will have some
positive eigenvalues and a similar discussion can go through. We note here that higher symmetries, e.g., a $Z_4$
symmetry, could lead to positive  eigenvalues of $Q_{mn}$ along more than one direction.
As a result additional directions would be unstable in the potential and  during
 inflation the brane can  roll along these additional directions as well.

{\it 4)  The Landscape and Inflation:}

{\it 4a) Broad Maxima:}
 
It seems quite likely that maximum of the kind found in this paper, which are broad with a small curvature,
are more generally present in the string theory landscape \cite{Susskind}. The unstable  directions
could be complex structure moduli or K\"ahler moduli as well. 
This possibility can be examined further as our understanding of moduli stabilization progresses. 
For complex structure moduli in the KKLT type constructions, one needs a better understanding of the 
non-perturbative corrections that are involved\footnote{In particular these contributions go like 
$W_{NP}=A(z_i) e^{-a\rho}$. The prefactor,  $A(z_i)$, depends on the complex structure moduli, and one 
 needs to  know  this dependence.}. As far as K\"ahler moduli are concerned, it seems difficult to 
get the required small mass with only one K\"ahler modulus,  since the canonically normalized field is 
logarithm of the volume.
But this might be possible with more than one K\"ahler moduli, again this requires further developments in our 
understanding of moduli stabilization. 

Generically broad maxima in the landscape, which agree with observational constraints,
 will have a  low scale of inflation\footnote{To see this, let us write the potential as follows,
\be
\label{genrpot} 
V=\Lambda^4 f(\phi/M_{Pl}) = \Lambda^4 - m^2 \phi^2 + {\cal{O}}(1) {\Lambda^4 \over M_{Pl}^4} \phi^4 + \cdots \,,
\ee
 where the ellipses denote higher powers of 
$(\phi/M_{Pl})$. Typically the slow roll conditions stop holding when the quartic terms come into play,
i.e., when  $\phi \sim M_{Pl} {m \over H} $.
Imposing $\eta ={m^2/3H^2} \sim 10^{-2}$, and the observed value of the density
 perturbations, eq.~(\ref{sdp}), then leads to $H \sim  10^{12} $ GeV. The resulting  power in 
gravity waves is ${\cal{P}}_{grav} \sim 10^{-14}$, which is 
quite small even for future observations. In specific models, like the one considered in 
this paper, inflation can end earlier and the resulting Hubble scale can be lower.} $H \lesssim 10^{12} {\rm GeV}$.
If inflation arises due to such a maximum in the landscape,  gravity waves are highly suppressed and will 
be difficult to observe in future experiments.

{\it 4b)  The Landscape and Various Inflationary Scenarios:}

So far in our discussion we have used the observational data and deduced various constraints on the  
compactification. Once explicit constructions become possible, we can turn this around and ask 
instead whether string theory makes any predictions about the observational data.  
While it is premature to speculate on this of course, it seems likely that in general the landscape will have 
many broad maxima of various different types, and the resulting values for the Hubble scale, $H$
and the mass, $m^2/H^2$, will take various different values, resulting in many different possibilities
for the scale of density perturbations and the tilt. Most of these will probably not agree with observation.

\bigskip
\goodbreak
\centerline{\bf Acknowledgments}
\noindent
We are grateful to Prasanta Tripathy for collaboration at early stages. 
S.P.T. acknowledges conversations with Alan Guth and other participants
 during the Cambridge workshop on Brane World
Cosmology. He also acknowledges discussions with participants of the String Theory and 
Cosmology workshop at KITP, Santa Barbara, and with the participants of the Coorg Conference 
on CMBR. N.I. thanks people at Harish-Chandra Research Institute in Allahabad 
for nice hospitality.  S.P.T. acknowledges support from the Swarnajayanti Fellowship,
D.S.T., Govt. of India. This research is supported by the generosity of the people of 
India, so we deeply thank  them.

%%%%%%%%%%%%%%%%%%%%%%%%%%%%%%%%%%%%%%%%%%%%%%%%%%%%%%%%%%%%%%%%%%%%%%%%%%%%%%%%

\end{document}